Menke, H.P.[1] (h.menke@hw.ac.uk)
Maes, J.[1] (j.maes@hw.ac.uk)
Geiger. S. (s.geiger@hw.ac.uk)

[1]Institute of GeoEnergy Engineering, Heriot-Watt University, Edinburgh, UK

Title: Upscaling the porosity-permeability relationship of a microporous carbonate to the Darcy scale with machine learning

# Upscaling the porosity-permeability relationship of a microporous carbonate to the Darcy scale with machine learning

Menke, H.P. [1]; Maes, J. [1]; Geiger, S. [1];

[1]Institute for GeoEnergy Engineering, Heriot-Watt University, Edinburgh, United Kingdom

## Abstract

The permeability of a pore structure is typically described by stochastic representations of its geometrical attributes (e.g. pore-size distribution, porosity, coordination number). Database-driven numerical solvers for large model domains can only accurately predict large-scale flow behavior when they incorporate upscaled descriptions of that structure. The upscaling is particularly challenging for rocks with multimodal porosity structures such as carbonates, where several different type of structures (e.g. micro-porosity, cavities, fractures) are interacting. It is the connectivity both *within* and *between* these fundamentally different structures that ultimately controls the porosity-permeability relationship at the larger length scales. Recent advances in machine learning techniques combined with both numerical modelling and informed structural analysis have allowed us to probe the relationship between structure and permeability much more deeply. We have used this integrated approach to tackle the challenge of upscaling multimodal and multiscale porous media.

We present a novel method for upscaling multimodal porosity-permeability relationships using machine learning based multivariate structural regression. A micro CT image of Estaillades limestone was divided into small $60^3$ and $120^3$ sub-volumes and permeability was computed using the Darcy-Brinkman-Stokes (DBS) model. The microporosity-porosity-permeability relationship from Menke et al. [1] was used to assign permeability values to the cells containing microporosity. Structural attributes (porosity, phase connectivity, volume fraction, etc.) of each sub-volume were extracted using image analysis tools and then regressed against the solved DBS permeability using an Extra-Trees regression model to derive an upscaled porosity-permeability relationship. Ten test cases of $360^3$ voxels were then modeled at the Darcy scale using this machine learning predicted upscaled porosity-permeability relationship and benchmarked against full DBS simulations, a numerically upscaled Darcy model, and a Kozeny-Carman model. All numerical simulations were performed using GeoChemFoam, our in-house open source pore-scale simulator based on OpenFOAM. We found good agreement between the full DBS simulations and both the numerical and machine learning upscaled models, with the machine learning model being 80 times less computationally expensive. The Kozeny-Carman model was a poor predictor of upscaled permeability in all cases.

## 1. Introduction

Predicting flow through porous media is pivotal for a broad range of scientific and engineering endeavors including fuel cells[2-4], oil and gas recovery[5-7], geologic carbon storage[8,9], geothermal energy[10-12], material composites[13], and nuclear waste disposal[14,15]. The internal structure of a porous media defines its ability to transmit fluid and therefore it's permeability. These porous structures are often heterogenous and range several orders of magnitude in scale, particularly in carbonate rocks which are an abundant geological material for both oil and gas reservoirs[16] and carbon storage sites[17-20], making prediction of permeability in these cases especially difficult.

Traditionally, permeability has been measured in core flood experiments using Darcy's law[21], which is based on bulk porosity measurements and the pressure drop across the core during flow. This method is expensive, time consuming and does not provide any insight into the local multi-scale structural influences on permeability, nor does it allow accurate extrapolation to other samples based on that structural information. Furthermore, flow measurements at the core scale are often not representative of flow at the reservoir scale due to large-scale heterogeneities present in the reservoir that are not captured in a small cm-scale core sample (e.g. vugs, fractures, or facies changes).

Recently, x-ray imaging has allowed us to image the physical heterogeneity of carbonate rocks at different scales: at the core (cm) scale using medical-CT imaging [22,23], at the pore (mm) scale using micro-CT imaging[24-26], and at the nano (μm) scale using nano-CT imaging[1,27-29]. These studies have given great insight into the types of structural heterogeneity seen at these distinctive scales. A choice few studies have attempted to bridge the gap between these scales[5,30]. Menke, et al. [1] applied correlative microscopy to incorporate nano-structural information of Estaillades microporosity imaging using nano-CT into pore-scale flow predictions using the Stokes-Brinkman equation. However, as of yet, no studies have integrated these structures into a single model to achieve accurate upscaling parameters that would allow information about the interactions between structures at multiple scales to inform Darcy-scale simulations.

Part of the difficulty in achieving robust upscaling parameters in carbonate rocks is analyzing the wealth of imaging data required to characterize the inherent rock heterogeneities at multiple scales. Over the past decade numerous open source computer vision tools have enabled in-depth analysis and vectorization of these immense datasets. New machine learning frameworks have also provided robust regression techniques that can facilitate variable prediction from multivariate statistics. Andrew [31] pioneered the combination of this data vectorization with the use of decision-tree based regression techniques to predict permeability from feature sets extracted from rock structures at a single scale. In this study we aim to expand on this technique both to: (1) predict the permeability of a multi-scale carbonate system using a combination of Darcy-Brinkman-Stokes (DBS) direct numerical simulation (DNS) approach and multivariate regression and then (2) test those machines learning derived upscaled permeability predictions in Darcy scale flow simulations and benchmark them against more computationally expensive DBS modeling methods, numerical upscaling, as well as the Kozeny-Carman (K-C) model, a popular simplistic upscaling technique that employs power-law trends in the porosity-permeability relationship.

First, a training set was created from a micro-CT image of Estaillades limestone that had been previously segmented into pore, solid grain, viton sleeve, and 12 phases of microporous grain based on their estimated voxel porosity[1]. The top 10% of the segmented micro-CT image was divided into 30,000 overlapping sub-volumes. 29,000 of the sub-volumes were solved for permeability using the DBS model GeoChemFoam[32] using the microporosity-porosity-permeability relationship from Menke, et al. [1] to assign permeability values to the Brinkman cells as a ground truth. Features of each sub-volume were extracted using image analysis tools (porosity, cumulative phase connectivity, and phase volume fraction). Using an Extra-Trees regression model[33], the extracted features were then regressed against the solved Brinkman permeability to achieve an upscaled permeability prediction model. 1000 sub-volumes not used in training were used to benchmark the regression model, finding a RMSE of 4.3% in permeability prediction. The remaining 90% of the micro-CT image not used in either the regression model or benchmarking was divided into ten $360^3$ voxel large-scale test volumes, which were then further subdivided into 6×6×6 and 3×3×3 matrices of sub-volumes of $60^3$ and $120^3$ voxels respectively. Features were extracted from each sub-volume and input into the trained machine learning regression model where the permeability of the sub-volumes was predicted. Darcy scale flow was then modelled on the test volumes using this regression-predicted upscaled permeability for each block and benchmarked against the full Stokes-Brinkman flow simulation of that block, the Darcy scale model using DBS numerical upscaling, as well as a Darcy scale Kozeny-Carman model fitted to the numerically predicted permeabilities. All numerical simulations were performed using GeoChemFoam, our open source in-house pore-scale simulator based on OpenFOAM [34,35].

## 2. Methods

### 2.1 Creating the Training Dataset

The porosity-permeability curve for the micro porosity is provided by the previous work in Menke, et al. [1]. A core of Estaillades limestone was scanned in a micro-CT both dry and saturated with a high contrast brine. The differential image was used to estimate the porosity of the connected micro porosity. Microporous subsections of the core were then scanned in a nano-CT and the grain size distribution of the micro porosity was modelled numerically and the results used to generate a synthetic porosity-permeability for the micro porosity. This curve was then used in a Stokes-Brinkman simulation of the whole core and benchmarked against experimental permeability measurements of the core with high accuracy. A detailed discussion of this multi-scale imaging and benchmark modelling can be found in Menke, et al. [1]. The image of Estaillades that is used in this study is the same as the image originally taken in Menke, et al. [1] and due to the extensive characterization performed in this study for the purposes of this work we will thus assume the permeability results and porosity-permeability curve derived in Menke, et al. [1] to be ground truth. The raw and processed micro and nano CT images are all available open access on the image archive of the British Geological Survey[36].

We used the 15-phase (pore, solid grain, viton sleeve and 12 phases of microporosity) segmented micro-CT image of Estaillades which is 1200x1200x6000 voxels with a resolution of 3.9 µm. The top 10% of the image was split into two training datasets: one with 30,000 sub-volumes of $60^3$ voxels with a 50% overlap between sequential sub-volumes to increase the number of training images and one with 30,000 sub-volumes of $120^3$ voxels with a 75% overlap between sequential sub-volumes. Overlapping the datasets allowed us to keep the two training datasets directly comparable.

Structural features were then extracted from each of the sub-volumes using the image analysis toolbox in python's SciKitImage[37]. This analysis included the volume fractions of each phase in the sub-volume as well as the cumulative connectivity of the phases in each orthogonal direction expressed mathematically as the first connected phase. An example of the calculation of the connectivity of the primary porosity of a sub-volume is shown in Figure 1. This image analysis created a robust feature vector set of 18 features: 15 volume fractions and 3 connectivity features.

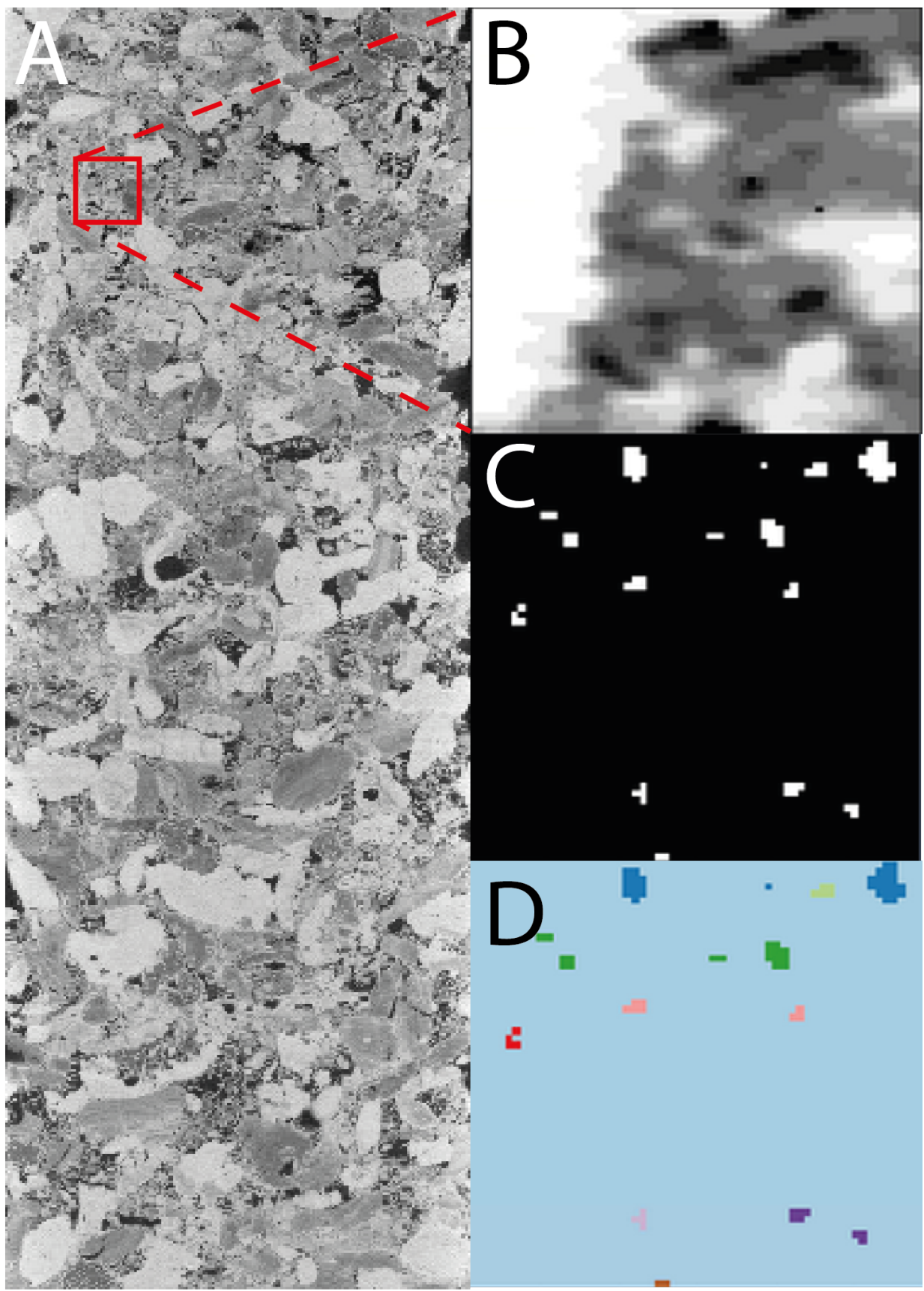

**Figure 1** Labelling of the connected components of the primary porosity in a sub-volume of the Estaillades sample. (A) The raw micro-CT image, (B) The 15-phase segmentation of a sub volume extracted from the microCT image (red box), (C) the segmented primary porosity, and (D) the uniquely labelled connected primary porosity.

## 2.2 Calculation of Permeability using GeoChemFoam

The training datasets (1) and (2) were solved for permeability in each direction using the DBS approach[32] in which one equation is used to model the flow within the fully resolved pores (i.e. voxel porosity equal to 1.0) and the micropores (i.e. voxel porosity lower than 1.0)

$$\frac{\mu}{\varphi}\nabla^2 u - \nabla p - \mu K^{-1} \cdot u = 0, \quad (1)$$

$$\nabla u = 0, \quad (2)$$

where $u$ [m.s-1] is the fluid velocity, $p$ [Pa] is the pressure, $\varphi$ [-] is the porosity, $\mu$ [kg.m$^{-1}$s$^{-1}$] is the viscosity and $K$ [m²] is the permeability. Porosity and permeability of each microporous voxel is assigned using the segmented phases and values from Menke, et al.[1] [Table 1].

**Table 1** Porosity and permeability values for micro-porosity used in the DBS simulations taken from Menke, et al.[1].

| Segmentation Phase # | Porosity [-] | Permeability [m²] |
|---|---|---|
| 1 | 1.00 | 1 (Pore) |
| 2 | 0.57 | 7.47 × 10$^{-15}$ |
| 3 | 0.52 | 6.91 × 10$^{-15}$ |
| 4 | 0.47 | 4.79 × 10$^{-15}$ |
| 5 | 0.42 | 3.24 × 10$^{-15}$ |
| 6 | 0.36 | 2.12 × 10$^{-15}$ |
| 7 | 0.27 | 8.06 × 10$^{-16}$ |
| 8 | 0.22 | 4.59 × 10$^{-16}$ |
| 9 | 0.18 | 2.44 × 10$^{-16}$ |
| 10 | 0.15 | 1.17 × 10$^{-16}$ |
| 11 | 0.12 | 4.95 × 10$^{-17}$ |
| 12 | 0.09 | 1.73 × 10$^{-17}$ |
| 13 | 0.07 | 4.76 × 10$^{-18}$ |
| 14 | 0.00 | 0 (Solid Grain) |
| 15 | 0.00 | 0 (Viton Sleeve) |

The model is implemented within GeoChemFoam. OpenFOAM solves Equations (1) and (2) on a collocated Eulerian grid. A pressure equation is obtained by injecting (1) into (2) and the system is solved using the Semi-Implicit Method for Pressure-Linked Equation (SIMPLE) algorithm[38].

For the upscaling models, Equation (1) can be simplified into the Darcy equations

$$-\nabla p - \mu K^{-1}.u = 0. \qquad (3)$$

Here $K$ is an anisotropic diagonal tensor that represents the permeability of each sub-volume ($60^3$ or $120^3$ voxels). Each diagonal coefficient ($K_x$, $K_y$ and $K_z$) has been computed using one of the three upscaling models presented in Section 3.4. Each sub-volume is modelled using a $4 \times 4 \times 4$ grid with constant coefficients, so that the total grid for the Darcy simulation is $24 \times 24 \times 24$ when using the $60^3$ sub-volumes and $12 \times 12 \times 12$ when using the $120^3$ sub-volumes.

## 3. Results & Discussion

The results and discussion are organized into four parts. (1) In section 3.1 we look at the relationship between porosity and permeability of the sub-volumes as solved by DBS and compare it to both, a power law fitted Kozeny-Carman model and the predicted permeabilities of our machine learning multivariate regression model. (2) In section 3.2 we examine the features used in the machine learning regression model and their importance relative to the number of features used in the regression as well as the error associated with the number of training images used to train the regression model. (3) In section 3.3 we compare the regression results from training the model with sub-volumes of different sizes. (4) Finally, in section 3.4 we use the trained regression models to upscale to the Darcy scale and compare this new machine learning based upscaling method with a brute force DBS simulation and both numerical and Kozeny-Carman upscaling models.

### 3.1 Sub-volume Permeability, Kozeny-Carman Model, and Multivariate Regression

The porosity and DBS solved permeabilities for training dataset 1 are shown in Figure 2A. A power law curve was used to estimate the Kozeny-Carman model parameters. The best model fit was $K = 8.47 \times 10^{-14} \frac{\varphi}{1-\varphi}^{3.4}$. Overall the Kozeny-Carman model was a poor fit for the training data as it does not capture the structural complexity inherent to the Estaillades carbonate where a single porosity value can result in permeabilities ranging over several orders of magnitude.

The feature vectors obtained from 29,000 of the 30,000 sub-volumes in training dataset 1 were then used to train an extra randomized trees ensemble machine learning regression model using SciKitLearn[33]. The remaining 1,000 sub-volumes were then used to test the regression model performance. The porosity values from these same sub-volumes were then plugged into the Kozeny-Carman model [Figure 2B]. The machine learning regression model significantly outperformed the Kozeny-Carman model where the RMSE of the Kozeny-Carman permeability predictions was 29.7% while the machine learning regression model had a RMSE of 4.3%. Furthermore, the Kozeny-Carman model's predicted values clustered towards the mean and the model was unable to predict the highest and lowest permeability values. This clustering indicates that the Kozeny-Carman model lacks enough complexity to incorporate the multiscale heterogeneities inherent in carbonate rocks with accuracy.

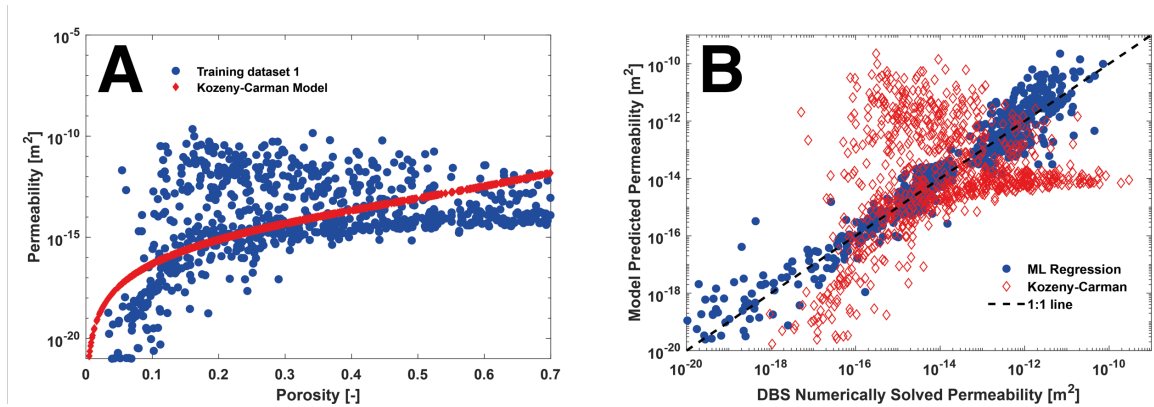

**Figure 2** (A) The porosity and numerically solved permeability for each of the 30,000 sub-volumes of $60^3$ voxels in training set 1 (blue) and the Kozeny-Carman fit (red) where $K = 8.47 \times 10^{-14} \frac{\varphi^{3.4}}{1-\varphi}$. (B) The machine learning regression model predicted permeability (blue) and the Kozeny-Carman permeability (red) plotted against the DBS solved permeability for 1,000 sub-volumes of $60^3$ voxels.

### 3.2 Analysis of Feature Importance

Three test cases of feature vectors, comprising 3, 15, and 18 features respectively, were assembled. In the '3 features' test set, only the connectivity feature vectors were included. In the '15 features' set, only the porosity volume fraction features vectors were considered, while in the '18 features' set, both the connectivity information and the porosity volume fraction features were included in the model. Figure 3A shows the RMSE of the different feature sets with the number of training sub-volumes used to train the model. All models showed a sharp drop in RMSE when increasing the number of sub-volumes used in the regression followed by diminishing returns as the number of sub-volumes increased past 500. This trend indicates that for this particular size of the sub-volume in this rock, very few sub-volumes are required to train an accurate model. However, this assertion is rock dependent, and we expect this number to change significantly with other rock types. Additionally, we found that the RMSE of the '18 features' case was the best with a minimum RMSE of 4.3%. However, this is only a slight improvement over the '3 features' set with a minimum RMSE of 5.6%, but a large improvement over the '15 features' set that has a minimum RMSE of 15.3%. This trend indicates that, as expected, the connectivity features are more important in predicting permeability than porosity alone.

Figure 3B shows the relative model weighting of each feature. Here we see that the connectivity features are weighted an order of magnitude more highly than the porosity features in all cases, confirming that when predicting permeability, the connectivity is a better predictor than porosity. This observation also highlights the importance of extracting the most informative features during the image processing and analysis. It would be interesting to also see if the connectivity could be quantified more robustly in the primary porosity (phase 1) to increase the accuracy of prediction in highly connected porosity sub-volumes. However, this analysis is out of the scope of this study and is a target of future work.

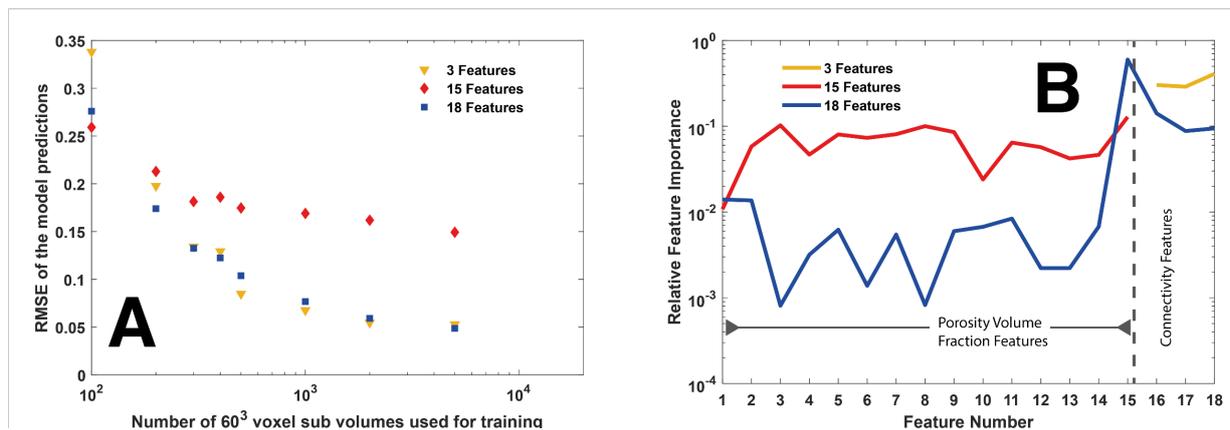

**Figure 3** (A) The RMSE of the of the different feature sets plotted against the number of training sub-volumes used to train the model. (B) The relative feature importance for regression models of training set 1 using 3 features (yellow), 15 feature (red), and 18 features (blue).

### 3.3 Analysis of the Size of the Training Data

As the ultimate goal of this work is to upscale to the Darcy scale, the choice of sub-volume size must be carefully considered. The sub-volumes must both be of a size relevant to Darcy scale imaging and modelling, but also be below the REV scale for porosity and permeability so that the spread of permeabilities is sufficient to capture structural heterogeneity while still containing enough tractable information on both the nano and micron scale porosity structures to make the feature selection in a multivariate regression representative of the characteristic properties influencing flow.

The image resolution of a medical CT scanner typically ranges between 250-500 microns, which corresponds approximately to the sizes of our sub-volumes chosen for investigation, i.e. $60^3$ voxels and $120^3$ voxels. To keep the total volume of rock in the training set constant (so as not to introduce any additional heterogeneity), the $120^3$ sub-volumes overlapped by 75% instead of the 50% in the $60^3$ sub-volumes. The same feature vector extraction workflow was used for both training datasets with between 100 and 29,000 training images and a further 1,000 test sub-volumes.

Figure 4A shows the RMSE of the regression model predictions against the number of sub-volumes used in the training for both the $60^3$ and $120^3$ cases. Both cases converge to approximately the same RMSE of ~4% with 5000 training images. However, from Figure 4B it is apparent that the spread of permeabilities is much greater for the $60^3$ case, indicating that the $60^3$ sub-volume size might be a better size to characterize the rock more accurately. This assertion is investigated further in Section 3.4.

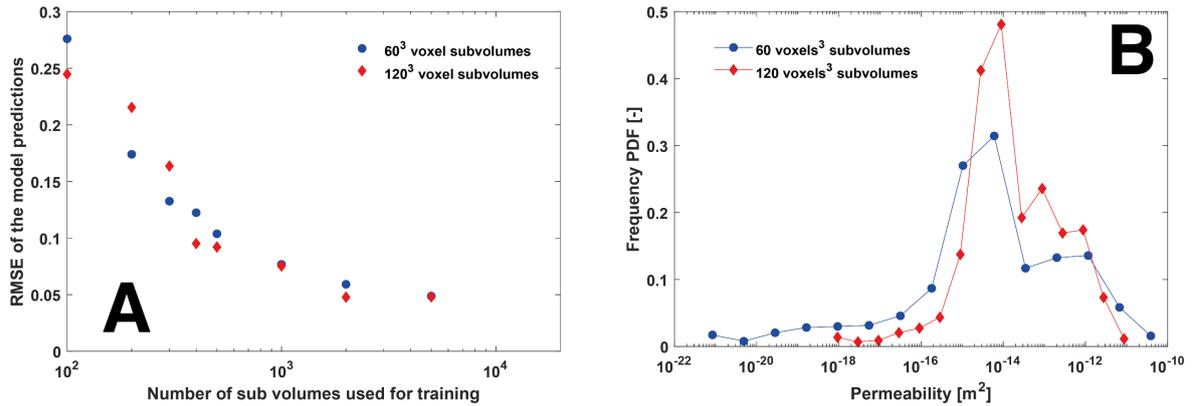

**Figure 4** (A) The RMSE of the model predictions with size of subvolume for both $60^3$ and $120^3$ subvolumes vs the number of subvolumes in the training dataset. (B) The PDF of ground truth (DBS) permeabilities for both $60^3$ and $120^3$ voxel sized sub-volume datasets.

### 3.4 Upscaling to the Darcy-scale

In this section we compare the ground truth DBS solution in ten $360^3$ test cases (Model 1) to the permeability predictions of three different upscaling models (Models 2,3, & 4). First, we divided the remaining 90% of the micro-CT image not used in regression model training or benchmarking into ten $360^3$ voxel test cases which were solved with DBS (Model 1). The test cases were then further subdivided into 6×6×6 and 3×3×3 matrices of sub-volumes of $60^3$ and $120^3$ voxels respectively for Darcy upscaling tests [Figure 5]. These sub-volumes were then solved with DBS and the output permeability used in each block of the Darcy model (Model 2). Next, the feature vectors were extracted for each of the sub-volumes and the trained regression model was used to predict the permeability of each sub-volume using the feature set. Each sub-volume was then assigned the permeability for the Darcy scale flow model with its calculated total porosity from the feature vector set (Model 3). Finally, the porosity of each sub-volume was used to predict permeability using the Kozeny-Carman model fit described in section 3.1 and that was used for each grid block of the Darcy scale model (Model 4).

The models are outlined as follows:

Model 1: Numerically solved $360^3$ volume with DBS (ground truth).

Model 2: Numerically solved both the $60^3$ and $120^3$ sub-volumes with DBS and used the output permeability to solve a Darcy simulation (numerical upscaling).

Model 3: Used the features of the $60^3$ and $120^3$ sub-volumes as input into the ML regression and then used the output permeability to solve a Darcy simulation (Machine Learning upscaling).

Model 4: Used the porosity of the $60^3$ and $120^3$ sub-volumes as input into the Kozeny-Carman model and then used the output permeability to solve a Darcy simulation (Kozeny-Carman upscaling).

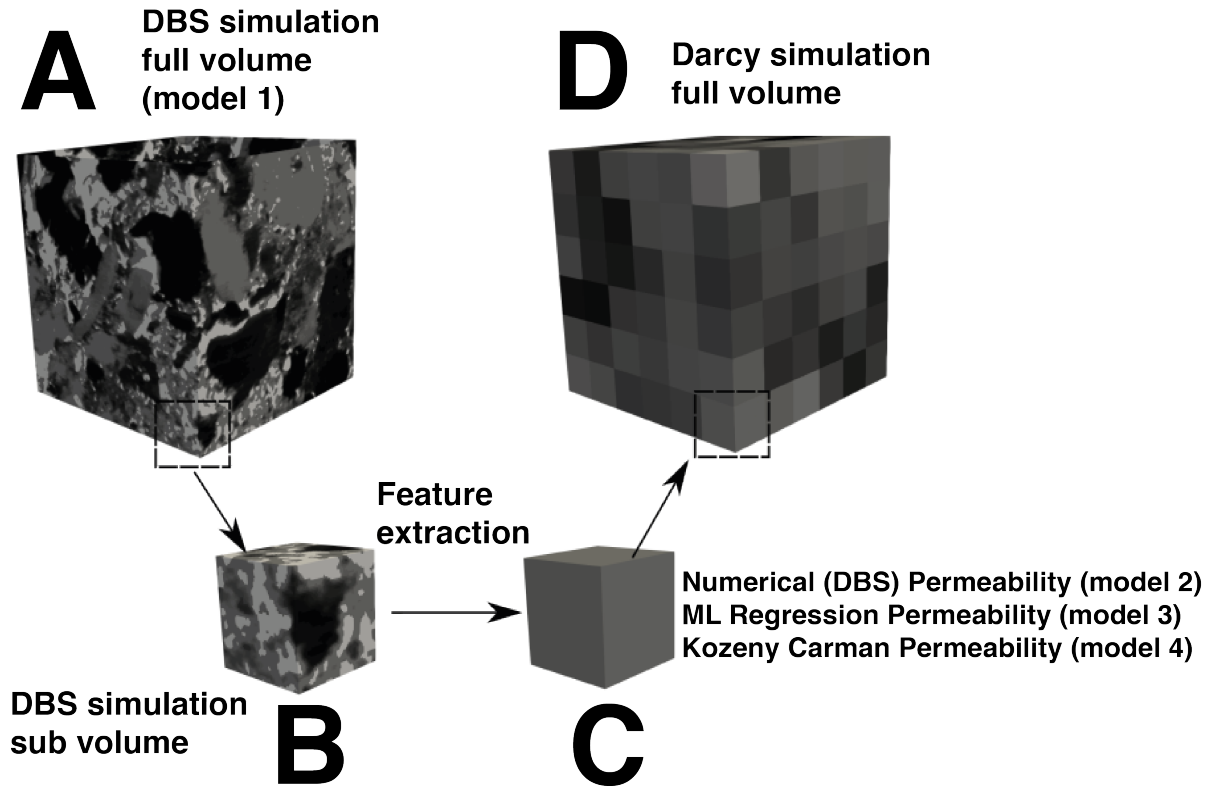

**Figure 5** The upscaling workflow: A $360^3$ voxel block (A) is cut into $60^3$ sub-volumes (B). The permeability for each sub-volume is then solved using DBS for numerical upscaling, the features extracted for machine learning upscaling, and the porosity calculated for Kozeny-Carman upscaling. The upscaled permeability for each model is then assigned to each upscaling Darcy block (C), and then the Darcy permeability is solved on the upscaled volume (D).

The permeability results of all these simulations are shown in Figure 6 and Table 2. We found that in the case of the $60^3$ sub-volumes the numerical upscaling and the machine learning upscaling both did equally well in predicting permeability with RMSE errors of 0.10 and 0.14, respectively, while the Kozeny-Carman did very poorly with a RMSE of 0.44. This poor performance is especially apparent in the case of test volume 2 which has the lowest porosity but the highest overall permeability in the ground truth, which is not something the Kozeny-Carman fit can predict. The upscaling results were not as accurate for the $120^3$ sub-volume cases with the machine learning upscaling slightly outperforming the numerical upscaling with a RMSE of 0.16 over the numerical upscaling's 0.11. The Kozeny-Carman predictions were the worst, with a RMSE of 0.58. The overall increase in error for the $120^3$ volumes can be explained by the smaller spread of sub-volume permeabilities shown in Figure 4B as compared to the $60^3$ sub-volumes. The $120^3$ sub-volumes are too big to characterize the range of flow heterogeneities in the rock at this resolution and thus the output permeabilities all trend towards the mean permeability. It is also important to note that in all cases the total CPU time for the machine learning model after model training was as low as the Kozeny-Carman model but had similar prediction performance to both, the full DBS case and numerical upscaling.

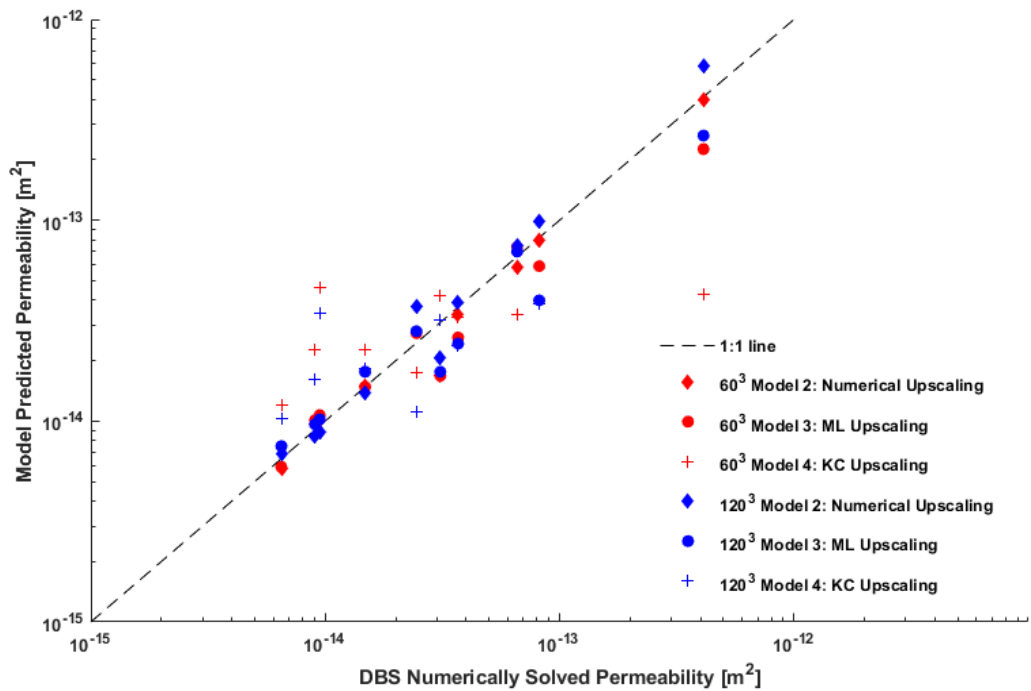

**Figure 6** The DBS calculated permeability of the $360^3$ voxel volumes plotted against the upscaled permeability results of Models 2, 3, and 4 using both, $60^3$ (red) and $120^3$ (blue) sub-volumes.

**Table 2** Results for $360^3$ Darcy blocks

| Size [voxels$^3$] | | Volume 1 | Volume 2 | Volume 3 | Volume 4 | Volume 5 | Volume 6 | Volume 7 | Volume 8 | Volume 9 | Volume 10 | RMSE | Total Run Time [min]* |
|---|---|---|---|---|---|---|---|---|---|---|---|---|---|
| 360 | Porosity | 0.36 | 0.43 | 0.35 | 0.40 | 0.38 | 0.34 | 0.40 | 0.36 | 0.34 | 0.34 | - | - |
| | Darcy Stokes Brinkman (ground truth) Permeability [m$^2$] | 6.59 × 10$^{-14}$ | 9.47 × 10$^{-15}$ | 7.67 × 10$^{-14}$ | 3.10 × 10$^{-14}$ | 3.69 × 10$^{-14}$ | 6.50 × 10$^{-15}$ | 8.19 × 10$^{-14}$ | 9.05 × 10$^{-15}$ | 2.45 × 10$^{-14}$ | 4.11 × 10$^{-13}$ | - | 1500 |
| 60 | Numerical DBS upscaled Permeability [m$^2$] | 5.85 × 10$^{-14}$ | 1.01 × 10$^{-14}$ | 1.50 × 10$^{-14}$ | 1.70 × 10$^{-14}$ | 3.40 × 10$^{-14}$ | 5.76 × 10$^{-15}$ | 7.93 × 10$^{-14}$ | 9.63 × 10$^{-14}$ | 3.71 × 10$^{-14}$ | 4.00 × 10$^{-13}$ | 0.11 | 240 |
| | Machine Learning upscaled Permeability [m$^2$] | 7.36 × 10$^{-14}$ | 1.07 × 10$^{-14}$ | 1.48 × 10$^{-14}$ | 1.68 × 10$^{-14}$ | 2.61 × 10$^{-14}$ | 5.91 × 10$^{-15}$ | 5.91 × 10$^{-14}$ | 1.01 × 10$^{-14}$ | 2.74 × 10$^{-14}$ | 2.26 × 10$^{-13}$ | 0.14 | 3 |
| | KC upscaled Permeability [m$^2$] | 3.40 × 10$^{-14}$ | 4.60 × 10$^{-14}$ | 2.26 × 10$^{-14}$ | 4.22 × 10$^{-14}$ | 3.30 × 10$^{-14}$ | 1.20 × 10$^{-14}$ | 3.97 × 10$^{-14}$ | 2.27 × 10$^{-14}$ | 1.74 × 10$^{-14}$ | 4.30 × 10$^{-14}$ | 0.44 | 3 |
| 120 | Numerical DBS upscaled Permeability [m$^2$] | 7.46 × 10$^{-14}$ | 8.79 × 10$^{-15}$ | 1.39 × 10$^{-14}$ | 2.08 × 10$^{-14}$ | 3.90 × 10$^{-14}$ | 6.83 × 10$^{-15}$ | 9.89 × 10$^{-14}$ | 8.39 × 10$^{-15}$ | 3.74 × 10$^{-14}$ | 5.89 × 10$^{-13}$ | 0.10 | 240 |
| | Machine Learning upscaled Permeability [m$^2$] | 6.98 × 10$^{-14}$ | 1.02 × 10$^{-14}$ | 1.76 × 10$^{-14}$ | 1.76 × 10$^{-14}$ | 2.43 × 10$^{-14}$ | 7.50 × 10$^{-15}$ | 3.99 × 10$^{-14}$ | 9.65 × 10$^{-15}$ | 2.80 × 10$^{-14}$ | 2.64 × 10$^{-13}$ | 0.16 | 3 |
| | Kozeny-Carman upscaled Permeability [m$^2$] | 1.49 × 10$^{-14}$ | 3.43 × 10$^{-14}$ | 1.83 × 10$^{-14}$ | 3.20 × 10$^{-14}$ | 2.37 × 10$^{-14}$ | 1.03 × 10$^{-14}$ | 3.84 × 10$^{-14}$ | 1.61 × 10$^{-14}$ | 1.12 × 10$^{-14}$ | 1.24 × 10$^{-14}$ | 0.58 | 3 |

*All model run times are for a 24 CPU workstation and summed across all volumes

# 4. Conclusion

In this study we have used the DBS model in GeoChemFoam in combination with decision tree based multivariate regression to upscale a microporous carbonate from the pore scale to the Darcy scale. We found that multivariate regression can be used to upscale and predict permeability with very few training images and performed equally well to numerical upscaling with a fraction of the computational cost. Additionally, increased sub-volume size had little effect on model predictions. However, a bigger sub-volume meant increased model CPU cost and decreased the accuracy of the upscaling models. Furthermore, we found that appropriate choice of feature vectors for extraction is important for regression model performance and connectivity information is the most important feature to include in the models. Machine learning based multivariate regression is thus an effective way of increasing prediction speed and accuracy during upscaling, but this method requires both precise multiscale imaging and an in depth understanding of connectivity in multiscale porosity structures. Future work will include investigation into using features that do not require high resolution imaging such as Darcy-scale porosity and tracer transport curves.

## Acknowledgements

This work was generously funded by the EPSRC project EP/P031307/1. SG also acknowledges Energi Simulation for supporting his chair program. Additionally, HPM would like to thank Dr Matthew G. Andrew for insightful conversations.